\documentstyle[twoside,fleqn,epsf,espcrc2]{article}

\newcommand{\AmS}{{\protect\the\textfont2
  A\kern-.1667em\lower.5ex\hbox{M}\kern-.125emS}}

\newcommand{\gsim}{{\protect
  \kern.18em\lower.5ex\hbox{$\stackrel{>}{\sim}$}\kern.25em}}
\newcommand{\lsim}{{\protect
  \kern.17em\lower.5ex\hbox{$\stackrel{<}{\sim}$}\kern.23em}}

\newcommand\be{\begin{equation}}
\newcommand\ee{\end{equation}}
\newcommand\bea{\begin{eqnarray}}
\newcommand\eea{\end{eqnarray}}

\title{
\begin{flushright}
  DESY 01-164
\end{flushright}
Overlap and domain wall fermions: 
what is the price of chirality? 
      }

\author{Karl Jansen \address{NIC/DESY Zeuthen,                                  
        Platanenallee 6, D-15738 Zeuthen, Germany}
       \thanks{e-mail: Karl.Jansen@desy.de} 
    }
           
\begin{document}

\begin{abstract}
In this contribution the costs of simulations employing  
domain wall and overlap fermions are estimated. 
In the discussion we will stay within the quenched approximation. 
\end{abstract}

\maketitle

Chirally invariant formulations of lattice QCD are still relatively 
new. However, besides their conceptual advantages,
a number of physics applications have already been performed
\cite{pilar_review}.
In these works 
it became clear that simulations with such formulations are
expensive, even when they are restricted to the quenched approximation.
Consequently, at the time of the conference no simulations
have been performed with such dynamical quarks so far. 

Our conclusions concerning the costs of simulations with dynamical
overlap or domain wall fermions will therefore be rather indirect:
we will estimate the overhead of using chiral invariant formulations
to standard Wilson fermions {\em in the quenched case} and assume
this overhead to be the same for an unquenched simulation. 
The discussion below will be organized in the form of four 
statements. 

\vspace{0.2cm}
\noindent{\bf Statement 1}
\vspace{0.2cm}

{\em The 5-dimensional domain wall construction in the limit
$N_s\rightarrow\infty$ is completely equivalent to a 4-dimensional
lattice formulation of overlap fermions.}

This is a mathematical statement that can be proven rigorously
\cite{shamir,furman,neuberger,noguchi,borici,hjl1}.
Denoting by $s$ the extra 5th dimension, the 
5-dimensional operator reads 

\be
D_5 = \frac{1}{2}\{\gamma_5(\partial_s^{*}+\partial_s)
               - a_s\partial_s^{*}\partial_s\} +M
\ee

with 
\be
M=D_w -m_0.
\ee
Here $D_w$ is the standard Wilson-Dirac operator, $a_s$ the
lattice spacing in the extra dimension and $m_0$ a mass parameter.

Keeping $a_s$ finite, in the limit $N_s\rightarrow\infty$ we obtain a
4-dimensional operator

\be
aD_4 = 1- A(A^\dagger A)^{-1/2}
\ee

with 
\be
A = -a_sM(2+a_sM)^{-1}
\ee

that satisfies the Ginsparg-Wilson relation. 
If also $a_s\rightarrow 0$,
Neuberger's overlap operator \cite{neuberger} is recovered.  

\vspace{0.2cm}
\noindent{\bf Statement 2}
\vspace{0.2cm}

{\em Domain wall fermions do not perform better than overlap 
fermions.}

The mathematical equivalence of domain wall and overlap fermions 
gives rise to the suspicion that also in practical applications 
no clear preference for either formulation --as far as the expense
of the simulation is concerned-- can be given. 
We give one example below for the cost of computing the pion 
propagator $\Gamma_{\pi}$ to a certain relative precision. 
In fig.~1 we plot the ratio $R$,

\be
R=\left| \frac{\Gamma_{\pi}^{\mathrm{exact}}(T/3)-\Gamma_{\pi}^{\mathrm{approx}}(T/3)}{
        \Gamma_{\pi}^{\mathrm{exact}}(T/3)}\right|\; .
\label{eq:theratioR}
\ee

Aiming at a, say, per mil precision we find for this particular 
example that domain wall fermions are about a factor of two more expensive
than overlap fermions. It is clear that for both formulations
additional improvements might be implemented. However, it seems
very unlikely that one particular formulation will give an 
order of magnitude better performance.

%%%%%%%%%%%%%%%%%%%%%%%%%%%%%FIGURE%%%%%%%%%%%%%%%%%%%%%%%%%%%%%%%%%%%
\vspace{-5mm}
\begin{figure}[htb] \label{figure1}
\centerline{ \epsfysize=7.8cm
             \epsfxsize=8.0cm
             \epsfbox{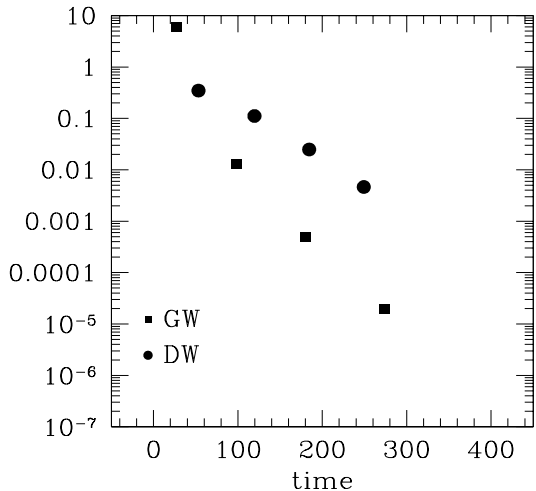}}
\vspace*{-0.9cm}
\caption{The ration $R$ of eq.~(\ref{eq:theratioR}) as a function of 
time needed (on a particular machine) to obtain a relativ 
precision for the pion propagator. Results are averaged over 
ten configurations
on a $8^3\times24$ lattice at $\beta=5.85$.}  
\end{figure}
%%%%%%%%%%%%%%%%%%%%%%%%%%%%%%%%%%%%%%%%%%%%%%%%%%%%%%%%%%%%%%%%%%%%%%

\vspace*{-0.6cm}
\noindent{\bf Statement 3}
\vspace{0.2cm}

{\em Keeping $N_s$ finite: the residual mass is not all.}

The locality of chirality breaking effects is only 
guaranteed at distances $|x-y|\gg N_sa$. This is discussed 
in \cite{hjl1,kikukawa}. First investigations of this question 
have been performed in \cite{christ}. I find it very important to 
study the effects of this observation further on a quantitative
level. 

\vspace{0.2cm}
\noindent {\bf Statement 4}
\vspace{0.2cm}

{\em Whatever time estimate is found for Wilson fermions:
multiply the effort by a factor of O(100) for chiral symmetric actions.}

The factor referred to in this statement is really only
an order of magnitude estimate. The reasons for this large factor 
are different for overlap and domain wall fermions. 

In the case of overlap fermions it turns out that the polynomial
required to approximate the square root has typically a degree
of O(100). Since this polynomial has to be evaluated in every 
step of a linear solver, the cost of simulations
with overlap fermions increase correspondingly. It is also noteworthy 
that so far no way of preconditioning the overlap operator 
has been found. 

In the case of domain wall fermions there is some experience \cite{hjl1}
that
the number of conjugate gradient iterations is larger for the 
5-dimensional problem. This in addition to the additional number of slices 
in the extra dimension gives again a large value of the factor 
compared to standard Wilson fermions. 

\vspace{0.2cm}
\noindent{\bf Conclusion}

If the estimate of about 100Tflops to solve most of the problems in
lattice QCD as discussed in the panel contributions is indeed correct,
then this would mean a demand of 10 Petaflops for simulations 
with chirally invariant
formulations of lattice-QCD. We then would be close to the number 
\cite{wilson} anticipated by K. Wilson 
in his 1989 Capri contribution.

When, as it is done at this conference, we emphasize that thinking 
about better algorithms is one of the most important things the
lattice community should address, then this is even more true 
for chirally invariant formulations of lattice-QCD. 
And, finally, the hope is that putting effort into the development of
better algorithms, it can be shown that statement number 4 is
not correct.

\end{document}